\documentclass[12pt]{article}
\usepackage{graphicx}
\usepackage{amsmath}
\usepackage{wrapfig}
\usepackage{amssymb}
\usepackage{lineno}
\usepackage{color}
\usepackage{relsize}

\def\babar {{{\mbox{\slshape B\kern-0.1em{\smaller A}\kern-0.1em B\kern-0.1em{\smaller A\kern-0.2em R }}}}}

\def\pbnr{}
\def\speaker{Valentina Santoro}
\def\onbehalfof{the \babar collaboration}
\def\title{Studies of Charmonium Production at \babar}
\def\affiliation{INFN Ferrara, via Saragat~1, 44122 Ferrara, Italy}
\def\support{The workshop was supported by the University of Manchester, IPPP, STFC, and IOP}

\newcommand\pubnumber{\pbnr}
\newcommand\pubdate{\today}
\def\Title#1{\begin{center} {\Large #1 } \end{center}}
\def\Author#1{\begin{center}{ \sc #1} \end{center}}

\newcommand{\OnBehalf}[1]{\sbox0{#1}\ifdim\wd0=0pt
        {}
	\else
	{\\on behalf of #1}
	\fi}
\newcommand{\SupportedBy}[1]{\sbox0{#1}\ifdim\wd0=0pt
        {}
	\else
	{\footnote{#1}}
	\fi}
\def\Address#1{\begin{center}{ \it #1} \end{center}}

\newcommand\pubblock{\includegraphics[width=5cm]{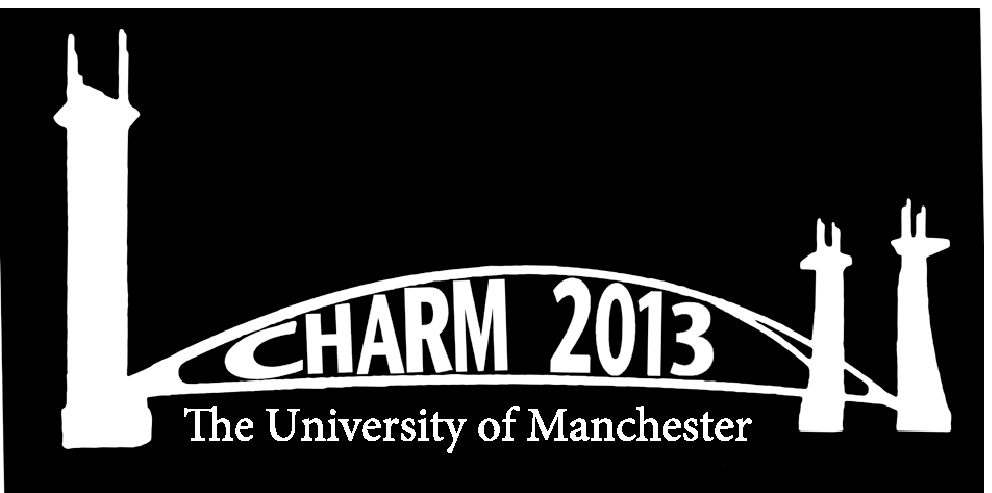}\hfill{\begin{tabular}{l} \pubnumber\\
         \pubdate  \end{tabular}}}
\newenvironment{Abstract}{\begin{quotation}  }{\end{quotation}}
\newenvironment{Presented}{\begin{quotation} \begin{center} 
             PRESENTED AT\end{center}\bigskip 
      \begin{center}\begin{large}}{\end{large}\end{center} \end{quotation}}

\def\venue{The 6$^{th}$ International Workshop on Charm Physics\\
(CHARM 2013)\\
Manchester, UK,  31 August -- 4 September, 2013}

\def\beq{\begin{equation}}
\def\eeq#1{\label{#1}\end{equation}}
\def\eeqn{\end{equation}}

\def\beqa{\begin{eqnarray}}
\def\eeqa#1{\label{#1}\end{eqnarray}}
\def\eeqan{\end{eqnarray}}

\let\bar=\overbar

\def\Dslash{\not{\hbox{\kern-4pt $D$}}}
\def\dslash{\not{\hbox{\kern-2pt $\del$}}}

\def\msb{{\bar{\ssstyle M \kern -1pt S}}}

\begin{document}
\begin{titlepage}
\pubblock

\vfill
\Title{\title}
\vfill
\Author{\speaker\SupportedBy{\support}\OnBehalf{\onbehalfof}}
\Address{\affiliation}
\vfill
\begin{Abstract}
We present recent results on charmonium and charmonium-like states from the \babar B-factory located at the PEP-II asymmetric-energy $e^{+}e^{-}$ collider at the SLAC National Accelerator Laboratory.
\end{Abstract}
\vfill
\begin{Presented}
\venue
\end{Presented}
\vfill
\end{titlepage}
\def\thefootnote{\fnsymbol{footnote}}
\setcounter{footnote}{0}
%


\section{Introduction}
The charmonium spectrum consists of eight narrow states below the open charm threshold (3.73~GeV) and several tens of states above that. 
Below the threshold almost all states are well-established. In contrast very little is known at higher masses where there have been discoveries~\cite{qr} of several new charmonium-like states for which the interpretation is still not clear. In the following sections we will review recent \babar results in this area.

\section{Study of $J/\psi \omega$ production in two-photon interactions}
The Y(3940) was observed by Belle \cite{Abe:2004zs} in B decays
 and confirmed by \babar  \cite{babary3940)}. In a re-analysis  \cite{X3872} of the \babar data sample the precision of the Y(3940) parameters was improved and evidence was found also for the decay $X(3872) \to J/\psi \omega$. This confirmed an earlier unpublished Belle claim  \cite{X3872belle}
 for the existence of this decay mode. A subsequent Belle paper \cite{X3915belle} reported evidence of a structure in the process $\gamma \gamma \to J/\psi \omega$ that they named the  X(3915), with mass and width values similar to those obtained for the Y(3940) by \babar \cite{babary3940)}. In this context \babar has performed a study of the process  $\gamma \gamma \to J/\psi \omega$~\cite{X3915BaBar}
 to search for the X(3915) and the X(3872) using a data sample corresponding to an integrated luminosity of 519 $\mathrm{fb^{-1}}$. 
 We searched for the X(3872) since, until recently~\cite{X3872lhcb}, its quantum numbers were ambiguous between $J^{PC}=1^{++}$ and $J^{PC}=2^{-+}$. For the former the state cannot be produced in two-photon collisions.
Figure 1 shows the reconstructed $J/\psi \omega$ mass distribution after all selection criteria have been applied. A large peak  near $3915~{\mathrm{MeV/c^{2}}}$ is observed with a significance of 7.6 $\sigma$. The measured parameters for the resonance, obtained from a maximum likelihood fit, are $m_{X(3915)}=(3919.4 \pm 2.2  \pm 1.6) {\mathrm {MeV/c^{2}}}$ and $\Gamma_{X(3915)}=(13\pm 6\pm 3)$MeV. The value of the two-photon width times the branching fraction is found to be $\Gamma_{\gamma \gamma}(X(3915)) \times {\cal B}(X(3915) \to J/\psi \omega) =52\pm 10 \pm 3$~eV for the spin 0 hypothesis, and $10.5 \pm 1.9 \pm 0.6$~eV for spin 2, where the first error is statistical and the second is systematic. 

\begin{figure}[htb]
  \centering
  \includegraphics[height=.24\textheight]{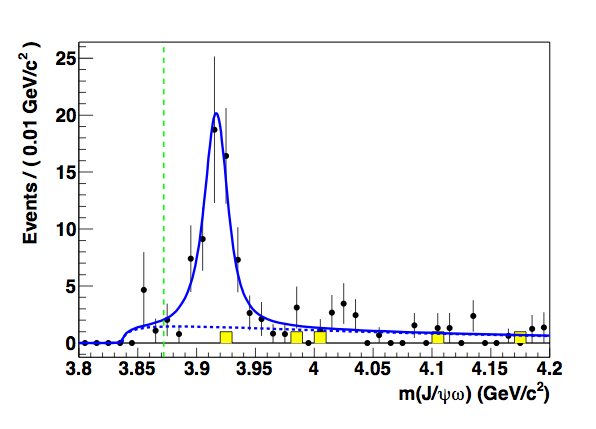}
  \label{fig1}
  \caption{The efficiency-corrected invariant mass distribution for the $J/\psi \omega$ final state. The solid curve represents the total fit function. The dashed curve is the background contribution. The shaded histogram is the non $J/\psi \omega$ background estimated from sidebands. The vertical dashed line is placed at the nominal X(3872) mass.}
\end{figure}
We performed an angular analysis based on the predictions of Rosner~\cite{Rosner:2004ac} in an attempt to establish the quantum numbers of the 
X(3915).   
We first discriminate between  $J^{P}=0^{\pm}$ and $J^{P}=2^{+}$ using the relevant final state angular distributions. In all cases
the $J^{P}=0^{\pm}$ hypothesis describes the data better than the $J^{P}=2^{+}$ hypothesis ~\cite{X3915BaBar}. We then discriminate between $J^{P}=0^{-}$ and $J^{P}=0^{+}$. In all cases the $J^{P}=0^{+}$ hypothesis gives a smaller $\chi^{2}$. In summary we find that assignment of  $J^{P}=0^{+}$ is preferred. This assignment favors the interpretation of the X(3915) as the $\chi_{c0}(2P)$ charmonium state.

\begin{figure}[htb]
  \centering
 \includegraphics[height=.4\textheight]{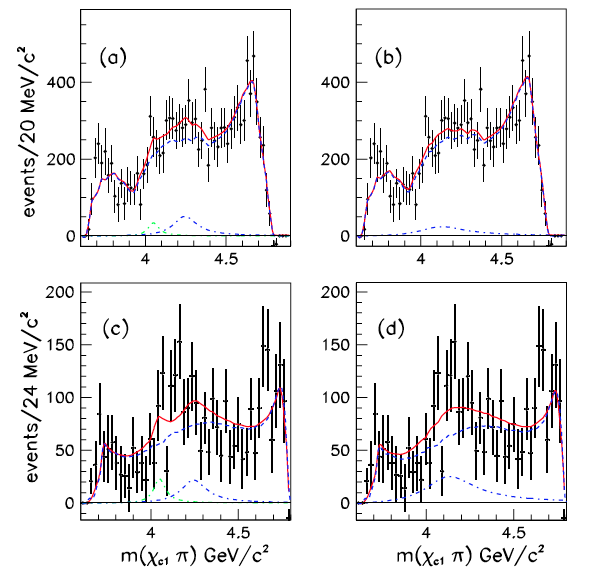}
  \label{fig4}
  \caption{(a),(b) Background-subtracted and efficiency-corrected $\chi_{c1}\pi$ mass distribution for $B\to \chi_{c1}K\pi$. (a) Fit with the $Z_{1}(4050)^{+}$ and $Z_{2}(4250)^{+}$ resonances. (b) Fit with only the $Z_{1}(4050)^{+}$ resonance. (c),(d) Efficiency-corrected and background-subtracted $\chi_{c1}\pi$ mass distribution in the $K\pi$ mass region for which Belle found the maximum resonance activity: $1.0~<~m^{2}(K\pi)~<~1.75~{\mathrm{ GeV^{2}/c^{4}}}$. (c) Fit with $Z_{1}(4050)^{+}$ and $Z_{2}(4250)^{+}$ resonances. (d) Fit with only the $Z(4150)^{+}$ resonance. The dot-dashed curves indicate the fitted resonant contributions.}
  \end{figure}
 
 \section{Search for the $Z_{1}(4050)^{+}$ and $Z_{2}(4250)^{+}$}
In 2008 the Belle Collaboration reported the observation of a resonance-like structure called the $Z(4430)^{+}$ decaying to $\psi(2S)\pi^{+}$ in the process $B\to \psi(2S)K\pi$ \cite{zbelle}. This claim generated a great deal of interest \cite{karlip} since such a state must have a minimum quark content $c\bar{c}\bar{d}u$, and thus would represent an unequivocal manifestation of a four-quark meson state. The \babar collaboration searched for the  $Z(4430)^{+}$ in an analysis of the process $B\to \psi(2S)K\pi$, and also in $B\to J/\psi K\pi$ \cite{zbabar},
but without finding significant structure in $\psi(2S) \pi$ nor in $J/\psi \pi$ invariant mass. In 2009 the Belle Collaboration reported the observation of two additional resonance-like structures similar to the $Z(4430)^{+}$ in the study of $\bar{B}^{0}\to \chi_{c1}K^{-}\pi^{+}$ \cite{z12belle}. These new structures were labeled as the $Z_{1}(4050)^{+}$ and the $Z_{2}(4250)^{+}$, both decaying to $\chi_{c1}\pi^{+}$.\\
\indent Using a data sample from an integrated luminosity of 429 $\mathrm{fb^{-1}}$, \babar has searched for the $Z_{1}(4050)^{+}$ and $Z_{2}(4250)^{+}$in the processes $\bar{B}^{0}\to \chi_{c1}K^{-}\pi^{+}$ and $B^{+}\to K_{s}^{0}\chi_{c1}\pi^{+}$ \cite{z12babar}, where the $\chi_{c1} \to J/\psi \gamma$. In the \babar analysis the $\chi_{c1}\pi^{+}$ mass distribution, after background subtraction and efficiency-correction, has been modeled using the angular information from the $K\pi$ mass distribution as represented using only low-order Legendre polynomial moments. 
The excellent description of the $\chi_{c1}\pi^{+}$ mass distribution obtained in this approach shows no need for any additional 
resonance structure in order to describe the distribution.  
Figure~\ref{fig4} shows the result of the fit to the $\chi_{c1}\pi^{+}$ mass spectrum using two or one scalar Breit-Wigners with parameters fixed to the Belle measured values. In all the fit cases there are no significant resonant structures, since the statistical significance obtained is less than $2 \sigma$. The upper limits (ULs) at the 90 \% CL on the branching fractions are, for the one resonance fit
${\cal B}(\bar{B}^{0} \to Z^{+}K^{-}) \times {\cal B}(Z^{+} \to \chi_{c1}\pi^{+})<4.7 \times 10^{-5}$,  while for the two-resonance fit 
${\cal B}(\bar{B}^{0} \to Z_{1}^{+}K^{-}) \times {\cal B}(Z^{+}_{1} \to \chi_{c1}\pi^{+})<1.8 \times 10^{-5}$ and ${\cal B}(\bar{B}^{0} \to Z^{+}_{2}K^{-}) \times {\cal B}(Z^{+}_{2} \to \chi_{c1}\pi^{+})<4.0 \times 10^{-5}$.

  \begin{figure}[htb]
  \centering
 \includegraphics[height=.2\textheight]{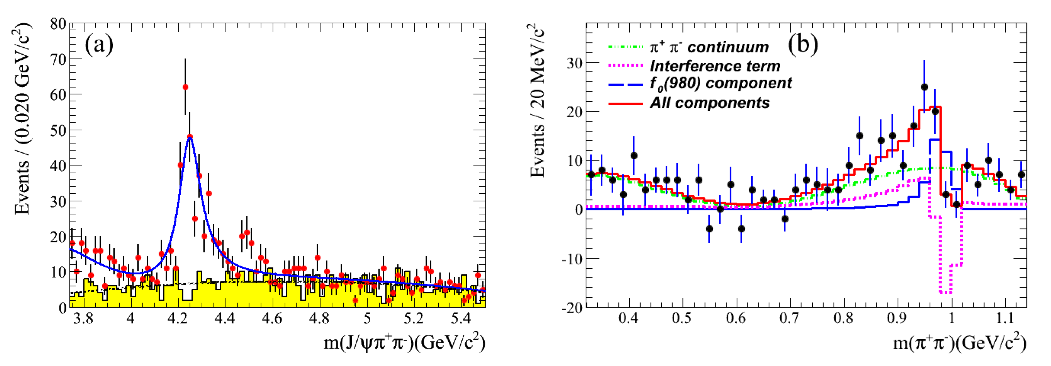}
 \label{fig3}
   \caption{(a): The $J/\psi \pi^{+}\pi^{-}$ mass spectrum from 3.74 ${\mathrm{GeV/c^{2}}}$ to 5.5 ${\mathrm{GeV/c^{2}}}$; the points represent the data and the shaded histogram is the background from the $J/\psi$~sidebands; the solid curve represents the fit result. (b) The $\pi^{+}\pi^{-}$ mass distribution from Y(4260) decay  to $J/\psi \pi^{+}\pi^{-}$. The solid histogram represents the result of the fit using the model described in the text.}
    \end{figure}

  \section{Study of the $J/\psi\pi^{+}\pi^{-}$ system via Initial State Radiation (ISR)}
In 2005 \babar discovered the Y(4260) in the process $e^{+}e^{-} \to \gamma_{ISR} Y(4260)$, with the $Y(4260) \to J/\psi\pi^{+}\pi^{-}$ \cite{Ybabar}.
Since this resonance is produced directly in $e^{+}e^{-}$ annihilation it has $J^{PC}=1^{--}$.
The 	observation of the decay
mode $J/\psi\pi^0\pi^0$ \cite{Ypi0} established that it has zero isospin. 
However it is not observed to decay to $D^*\bar{D^*}$ \cite{Ydd}, nor to $D_s^*\bar{D^*_s}$ \cite{Yds}, so that its properties do not lend 
themselves to a simple charmonium interpretation, and its nature remains unclear.
A subsequent Belle analysis \cite{Ybelle} of the same final state suggested also the existence
of an additional resonance around 4.1 GeV/c$^2$ that they named the Y(4008).
\babar has performed  an analysis \cite{Ybabarnew}
 of this process using a data sample corresponding to an integrated luminosity of 454 ${\mathrm fb^{-1}}$.  Figure~\ref{fig3}(a) shows the invariant mass distribution for $J/\psi\pi^{+}\pi^{-}$ after all selection criteria have been applied . A clear signal for the Y(4260) is seen. We performed an unbinned-maximum-likelihood fit, and obtained $m_{Y(4260)}=4244 \pm 5 \pm4~{\mathrm {MeV/c^{2}}}$, $\Gamma_{Y(4260)}=114^{+16}_{-15} \pm 7$ MeV and $\Gamma_{ee} \times {\cal B}(J/\psi \pi^{+}\pi^{-}) =9.2 \pm 0.8 \pm 0.7$~eV. There is no evidence for the Y(4008) found by Belle \cite{Ybelle}. In this \babar analysis a detailed study of the $\pi^{+}\pi^{-}$ system from the Y(4260) decay  to $J/\psi \pi^{+}\pi^{-}$ has been performed. The $\pi^{+}\pi^{-}$ mass distribution shown in Figure 3(b) peaks near the $f_{0}(980)$ mass, but is displaced from the nominal $f_{0}(980)$ position, and occurs at $\sim$ 940 ${\mathrm{MeV/c^{2}}}$. The fact that the peak is displaced, together with the particular shape of $m(\pi^{+}\pi^{-})$ distribution, suggests the possibility interference between the $f_{0}(980)$ and an $m(\pi^{+}\pi^{-})$ continuum. To test this possibility the $f_{0}(980)$ amplitude and phase have been taken from the \babar analysis \cite{antimo} of  $D_{s}^{+}\to \pi^{+}\pi^{-}\pi^{+}$ and this complex amplitude has been used in a simple model to describe the $\pi^{+}\pi^{-}$ mass distribution of the form $|\sqrt{pol} +e^{i\phi}F_{f_{0}(980)}|^{2}$ where $``pol"$ is a polynomial function used to describe the  $m(\pi^{+}\pi^{-})$ continuum, and $F_{f_{0}(980)}$ is the amplitude from $D_{s}^{+}\to \pi^{+}\pi^{-}\pi^{+}$ \cite{antimo} analysis; $\phi$ allows for a phase difference between these amplitudes. The result of this fit is shown in Figure~\ref{fig3}(b) and it indicates that if there is a real $f_{0}(980)$ contribution to the decay of the Y(4260) to~$J/\psi\pi^{+}\pi^{-}$ its contribution is small, since we obtain $\frac{{\cal B}(Y_{4260} \to J/\psi f_{0}(980), ~f_{0}(980)\to \pi^{+}\pi^{-})}{{\cal B}(Y_{4260}   \to J/\psi \pi^{+}\pi^{-})}=(17 \pm 13) \%$.

\section{Conclusion}
We have presented studies of charmonium-like states at \babar. We have confirmed the existence of the X(3915), and determined its preferred quantum numbers to be  $J^{P}=0^{+}$. We also presented the search for the $Z_{1}(4050)^{+}$ and $Z_{2}(4250)^{+}$, and the update of the \babar analysis of the decay $Y(4260) \to J/\psi \pi^{+}\pi^{-}$. All these measurements may help our understanding of the charmonium-like states discovered at the B-factories.

\end{document}